\documentclass[aps,pre,twocolumn,showpacs]{revtex4-1}
\usepackage{graphicx,amsbsy,amsmath,epsfig,natbib,float}
\usepackage{amssymb,latexsym,wasysym,pifont,mathrsfs}
\usepackage{comment,verbatim,multirow,array,sidecap,color}
\usepackage{lineno,setspace,soul,cancel}
\usepackage[normalem]{ulem}

\begin{document}

\title{Dissipative stochastic sandpile model on small world network :
  properties of non-dissipative and dissipative avalanches}

\author{Himangsu Bhaumik}
\email{himangsu@iitg.ac.in}
\author{S. B. Santra}
\email{santra@iitg.ac.in}
\affiliation{Department of Physics, Indian Institute of Technology
  Guwahati, Guwahati-781039, Assam, India.}

\date{\today}
 
\begin{abstract}  
 A dissipative stochastic sandpile model is constructed and studied on
 small world networks in one and two dimensions with different
 shortcut densities $\phi$, where $\phi=0$ represents regular lattice
 and $\phi=1$ represents random network. The effect of dimension,
 network topology and specific dissipation mode (bulk or boundary) on
 the the steady state critical properties of non-dissipative and
 dissipative avalanches along with all avalanches are analyzed. Though
 the distributions of all avalanches and non-dissipative avalanches
 display stochastic scaling at $\phi=0$ and mean-field scaling at
 $\phi=1$, the dissipative avalanches display non trivial critical
 properties at $\phi=0$ and $1$ in both one and two dimensions. In the
 small world regime ($2^{-12} \le \phi \le 0.1$), the size
 distributions of different types of avalanches are found to exhibit
 more than one power law scaling with different scaling exponents
 around a crossover toppling size $s_c$. Stochastic scaling is found
 to occur for $s<s_c$ and the mean-field scaling is found to occur for
 $s>s_c$. As different scaling forms are found to coexist in a single
 probability distribution, a coexistence scaling theory on small world
 network is developed and numerically verified.
\end{abstract}

\maketitle

\section{Introduction} 
Power-law scaling in many natural phenomena such as earthquakes
\cite{chenPRA90}, forest fires \cite{drosselPRL92,*drosselPRL93},
biological evolution \cite{bakPRL93}, droplet formation
\cite{plourdePRL93}, superconducting avalanches \cite{fieldPRL95},
etc. are found to be outcome of self organized criticality (SOC)
\cite{bak,*jensen,*christensen-moloney,*pruessner} which refers to the
intrinsic tendency of a wide class of slowly driven systems to evolve
spontaneously to a non equilibrium steady state. At the same time,
self-organization on complex structures or networks are found to
appear very often in nature. For example, avalanche mode of activity
in the neural network of brain \cite{arcangelisPRL06,*hesseFSN14},
earthquake dynamics on the network of faults in the crust of the earth
\cite{lisePRL02}, rapid rearrangement of coronal magnetic field
network \cite{hughesPRL03}, propagation of information through a
network with a malfunctioning router causing the breakdown of the
Internet network \cite{motterPRE02}, blackout of the electric power
grid \cite{carrerasIEEECS04} and many others. On the other hand, small
world network (SWN) \cite{wattsNAT98} not only interpolates between
the regular lattice and the random network but also it preserves both
the properties of regular lattice and random network, namely high
``clustering-coefficient''(concept of neighborhood) and
``small-world-effect'' (small average shortest distance between any
two nodes) respectively. It is always intriguing to study the models
of SOC on networks such as SWN.

Sandpile, a prototypical model to study SOC introduced by Bak, Tang,
and Wiesenfeld (BTW) \cite{btwPRL87,*btwPRA88}. Though BTW on regular
lattice gives rise to anomalous (multi) scaling
\cite{deMenechPRE98,*tebaldiPRL99,lubeckPRE00a}, it shows a mean-field
scaling \cite{christensenPRE93,bonabeau95,gohPRL03,bhaumikPRE13} when
studied on random network. A transition from non-critical to critical
behaviour was reported in BTW type sandpile model on SWN in one
dimension ($1$d) \cite{ lahtinenPHYA05} whereas a continuous crossover
to mean-field behaviour was reported for the same model on SWN in two
dimension ($2$d) \cite{arcangelisPHYA02}. However recent study of BTW
model on SWN in $2$d shows the co-existence of more than one scaling
forms in the distributions of avalanche properties
\cite{bhaumikPRE13}. On the other hand, stochastic sandpile models
(SSM) on regular lattice, which incorporates random distribution of
sand grains during avalanche, exhibit a scaling behaviour with
definite critical exponents that follows finite size scaling (FSS) and
define a robust universality class called Manna class
\cite{mannaJPA91,*dharPHYA99a}. More insight in avalanche size
distribution statistics were obtained by classifying the avalanches
into dissipative and non-dissipative avalanches. The size distribution
of dissipative avalanches of BTW sandpile in $2$d was found to follow
power law scaling with a definite exponent that does not obey FSS
\cite{drosselPRE00}. Later, Dickman and Campelo \cite{dickmanPRE03}
showed both in one and two dimensions that the dissipative and
non-dissipative avalanches of SSM on regular lattice obey different
FSS behaviour with certain logarithmic correction beside the power law
scaling with different exponents. However, there are not many studies
that report the critical behaviour of SSM and specially that of the
stochastic dissipative avalanches on networks. It is then intriguing
to study avalanche size distribution of dissipative and
non-dissipative avalanches of a stochastic sandpile model on SWN which
interpolates regular lattice and random network, and verify whether
all such scaling forms would be preserved or not under bulk
dissipation mode.

In this paper, a dissipative stochastic sandpile model (DSSM) is
constructed on SWN and studied as a function of shortcut density
$\phi$ in both one and two dimensions. The distribution functions of
the steady state avalanche properties as well as those of dissipative
and non-dissipative avalanches on regular lattice ($\phi=0$) and
random network ($\phi=1$) are found to display several interesting
non-trivial features which are not reported before. Moreover, in the
small world regime with intermediate $\phi$ ($\approx$ $2^{-12}$ to
$2^{-3}$) \cite{newmanbook}, the steady state avalanche properties
exhibit coexistence of the SSM scaling and the mean-field scaling in a
single distribution depending on the avalanche sizes. A coexistence
scaling theory is developed and numerically verified.

\section{The Model} 
SWN is generated both on a $1$d linear lattice and on a $2$d square
lattice by adding shortcuts between any two randomly chosen lattice
sites which will be referred as nodes later on. The shortcut density
$\phi$ is defined as the number of added shortcuts $N_\phi$ per
existing bond ($dL^d$ bonds are present in a $d$-dimensional lattice
of linear size $L$ with periodic boundary conditions (PBC) and without
shortcuts) and is given by $\phi=N_\phi/(dL^d)$. Care has been taken
to avoid self-edges of any node and multi-edges between any two
nodes. To study sandpile dynamics on an SWN, first an SWN is generated
for a particular value of $\phi$ and it is then driven by adding sand
grains, one at a time, to randomly chosen nodes. If the height $h_i$
of the sand column at the $i$th node becomes greater than or equal to
the predefined threshold value $h_c$, which is equal to $2$ here, the
$i$th node topples and the height of the sand column of the $i$th node
will be reduced by $h_c$. The sand grains toppled are then distributed
among two of its randomly selected adjacent nodes which are connected
to the toppled node either by shortcuts or by nearest neighbour
bonds. During distribution of the sand grains PBC is applied. Hence,
there is no open boundary in the system where dissipation of sand
grains could occur. A dissipation factor $\epsilon_\phi$ is then
introduced during transport of a sand grain from one node to another
to avoid overloading of the system. The toppling rule of $i$th
critical node in this DSSM on SWN then can be represented as
\begin{equation}
\label{trule2}
\begin{array}{ll}
& h_i \rightarrow h_i-h_c, \\ {\rm and} & h_j=
  \left\{\begin{array}{ll} h_j + 0 & \mbox{if} \hspace{0.2cm} r \le
  \epsilon_\phi, \\ h_j + 1 & \mbox{otherwise} \end{array}\right.
\end{array}
\end{equation}
where $j$ is two randomly selected nodes out of $k_i$ adjacent nodes
of the $i$th node, $r$ is a random number uniformly distributed over
$[0,1]$. In this distribution rule, an adjacent node may receive both
the sand grains. If the toppling of a node causes some of the adjacent
nodes unstable, subsequent toppling follows on these unstable
nodes. The process continues until there is no unstable node present
in the system. These toppling activities lead to an avalanche. During
an avalanche no sand grain is added to the system.

For a given SWN, $\epsilon_\phi$ is taken as $1/\langle n_{\phi}
\rangle$, where $\langle n_{\phi} \rangle$ is the average number of
steps required for a random walker to reach the lattice boundary
(without PBC) starting from an arbitrary lattice site. There exits a
characteristic length $\xi\sim \phi^{-1/d}$ where $d$ is the
dimensionality of the lattice, below which SWN belongs to the ``large
world'', the regular lattice regime, and beyond which it behaves as
``small world'', the random network regime
\cite{newmanPRE99,*newmanPLA99,mendesEPL00}. The asymptotic behaviour
of $\langle n_{\phi} \rangle$ with $\phi$ and $L$ is given by
\begin{eqnarray}
\langle n_{\phi} \rangle&\sim& \left \{ \begin{array}{ll}
  L^2, & \phi \rightarrow 0\\
  L\phi^{-1/d}, & \phi \rightarrow 1\\
       \end{array} \right.
\label{nLp}
\end{eqnarray}
for a $d$ dimensional SWN. It has diffusive behaviour for $\phi
\rightarrow 0$ and super-diffusive behaviour for $\phi \rightarrow
1$. The above scaling form is numerically verified in
Ref. \cite{bhaumikPRE13}. The dissipation factor
$\epsilon_\phi=1/\langle n_{\phi} \rangle$ for a given $\phi$ is
determined using numerically estimated values of $\langle n_{\phi}
\rangle$. A few values of $\epsilon_\phi$ are listed in
Table. \ref{table} for $1$d and $2$d lattices.

\begin{table}[t]
\centering
\begin{tabular}{c@{\hspace{1cm}}c@{\hspace{0.1cm}}c@{\hspace{1cm}}c} 
  \hline\hline \multicolumn{1}{c}{\multirow{2}{*}{ $\phi$ }} &  \multicolumn{3}{c}{$\epsilon_\phi$}\\
  \cline{2-4} 
  & $d=1,L=8192$ && $d=2,L=1024$ \\\hline 
  \multicolumn{1}{c}{ $0$}     & $8.94\times10^{-8}$& & $6.83\times10^{-6}$  \\
  \multicolumn{1}{c}{ $2^{-9}$} & $5.10\times10^{-7}$& & $6.34\times10^{-5}$  \\
  \multicolumn{1}{c}{ $2^{-8}$} & $9.72\times10^{-7}$& & $9.12\times10^{-5}$  \\
  \multicolumn{1}{c}{ $2^{-7}$} & $1.81\times10^{-6}$& & $1.34\times10^{-4}$  \\
  \multicolumn{1}{c}{ $2^{-6}$} & $3.60\times10^{-6}$& & $1.99\times10^{-4}$  \\
  \multicolumn{1}{c}{ $2^{-5}$} & $7.10\times10^{-6}$& & $2.96\times10^{-4}$  \\
  \multicolumn{1}{c}{ $1$}     & $1.27\times10^{-4}$& & $2.06\times10^{-3}$  \\
  \hline\hline
\end{tabular}
\caption{\label{table} Dissipation factor $\epsilon_\phi$ for selected
  values of $\phi$ on $1$d lattice of $L=8192$ and $2$d square lattice
  of size $L=1024$.}
\end{table}

\section{Results and discussion}
Extensive computer simulations are performed to study the dynamics of
DSSM on SWN in $1$d and $2$d. After a transient period, the system
evolves to a steady state which corresponds to equal currents of sand
influx and outflux resulting constant average height of the sand
columns. Critical properties of DSSM on SWN are characterized studying
various avalanche properties in the steady state at different values
of $\phi$ and system size $L$. The maximum lattice size used for $1$d
is $L=8192$ and that for $2$d is $L=1024$. Data are averaged over
$32\times 10^6$ avalanches collected on $32$ different SWN
configurations for a given $\phi$ and $L$. The information of an
avalanche is kept by storing the number of toppling of every node in
an array $S_\phi[i],i=1,\cdots,L^d$ which was set to zero
initially. All geometrical properties of an avalanche such as
avalanche size $s$, avalanche area $a$, etc. can be estimated in terms
of $S_\phi[i]$ as given below
\begin{equation}
s=\sum_{i=1}^{L^d} S_\phi[i], \ \ \ a= \sum_{i=1}^{L^d}1
\end{equation}
for all $S_\phi[i]\ne 0$.

\subsection{Toppling Surface}
\begin{figure}[t]
\centerline{\hfill
  \psfig{file=figure_1ab.eps,width=0.48\textwidth}\hfill
}
\vspace{.5cm} \centerline{\hfill
  \psfig{file=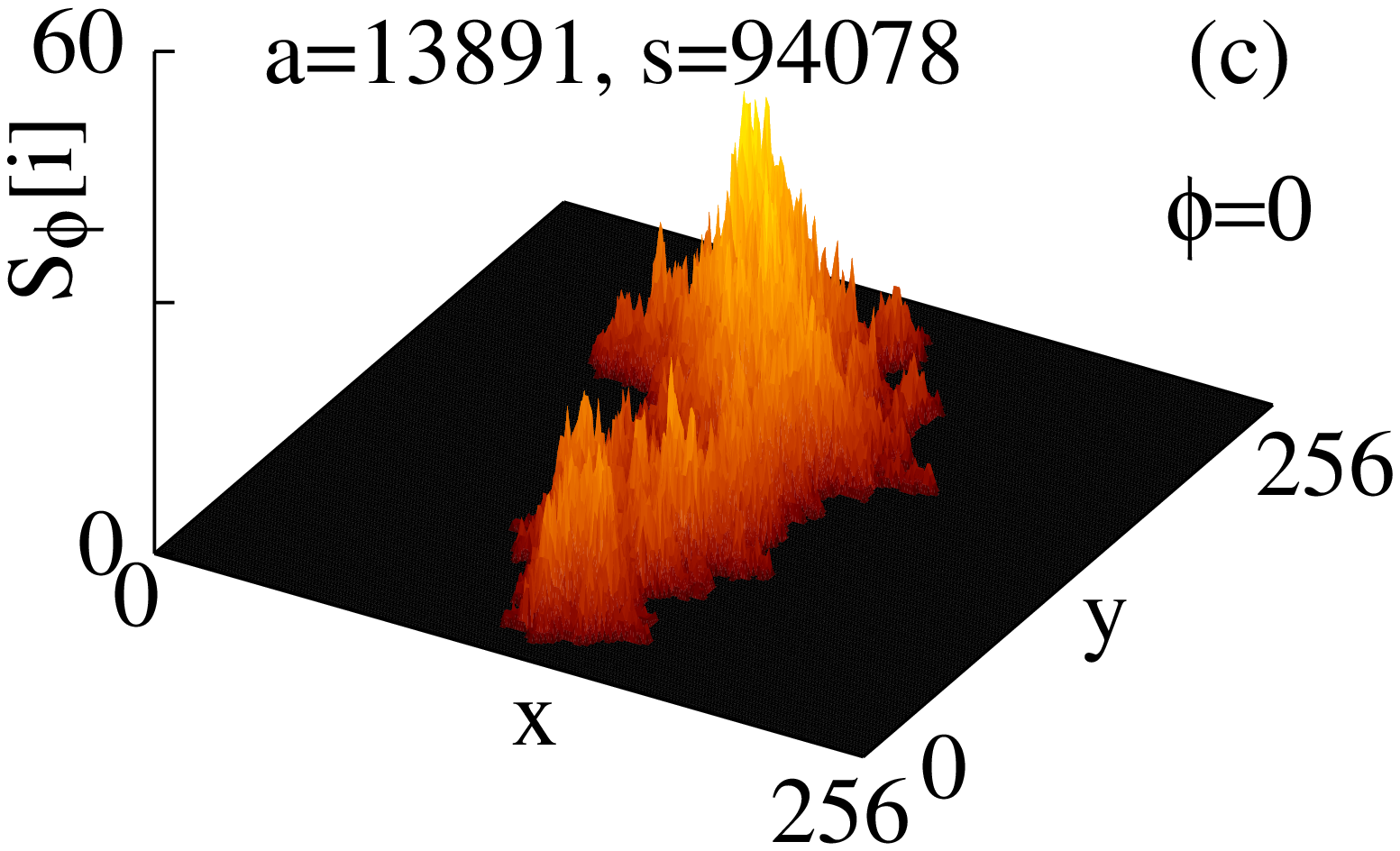,width=0.23\textwidth}\hfill
  \psfig{file=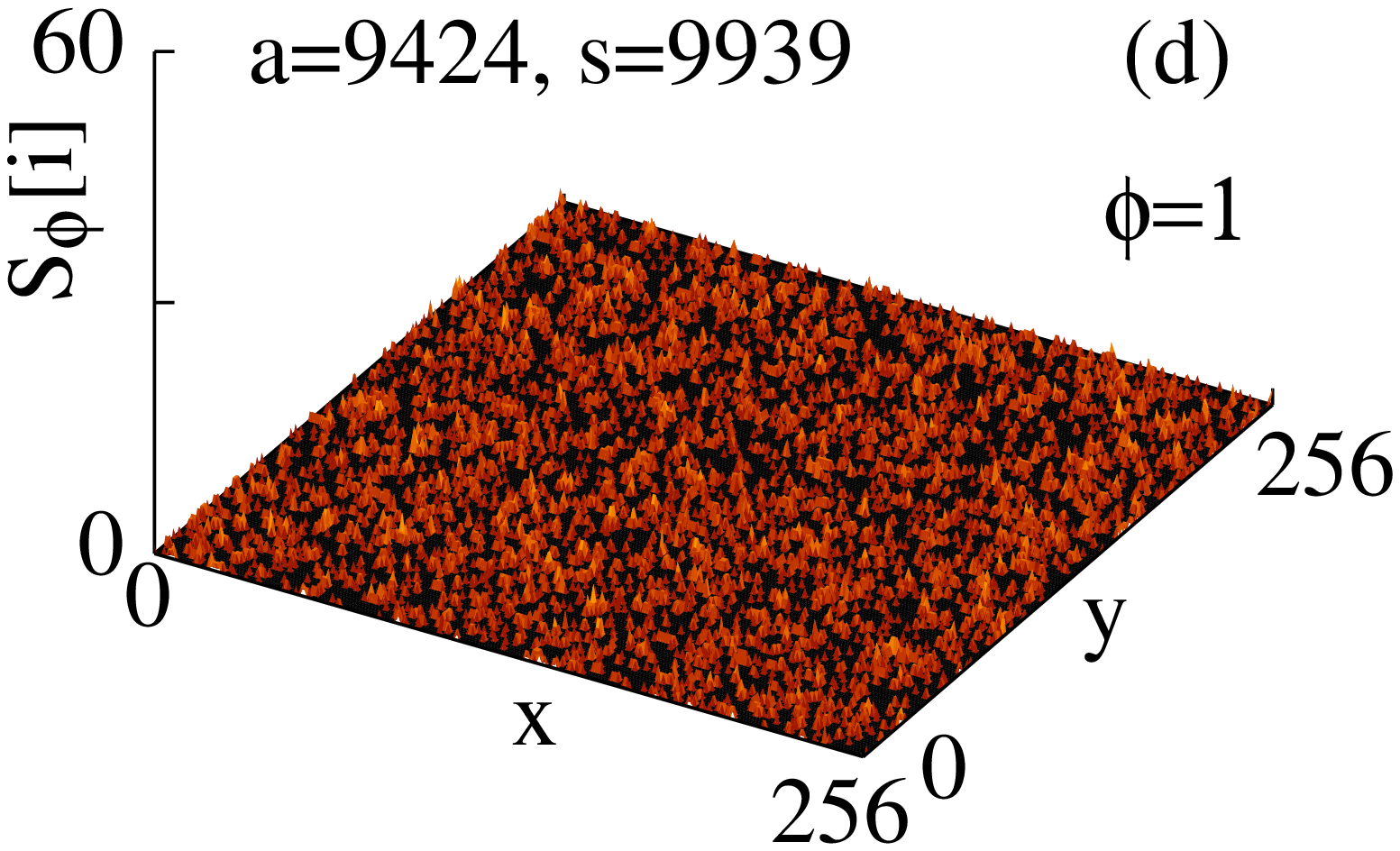,width=0.23\textwidth}\hfill }
\caption{(Color online) The toppling surfaces of typical large
  avalanches of DSSM on SWN are shown. Toppling surface of an
  avalanche generated on a $1$d lattice of size $L=256$ for $\phi=0$
  is in (a) and for $\phi=1$ is in (b). The surface generated on a
  $2$d square lattice of size $L=256$ for $\phi=0$ is in (c) and for
  $\phi=1$ is in (d). The size ($s$) and area ($a$) of the
  corresponding avalanches are mentioned as legend in the respective
  plots. }
\label{surf01}
\end{figure}
The values of the toppling number $S_\phi[i]$ of an avalanche at
different nodes of SWN define a surface called toppling surface
\cite{ahmedEPL10} which serves as an important geometrical quantity to
visualize an avalanche. The toppling surfaces for typical large
avalanches in the steady state, generated on a $1$d lattice of size
$L=256$, are presented for $\phi=0$ and $\phi=1$ respectively in
Figs. \ref{surf01}(a) and \ref{surf01}(b). Toppling surfaces generated
on a $2$d square lattice of size $L=256$ are presented in
Fig. \ref{surf01}(c) for $\phi=0$ and for $\phi=1$ in
Fig. \ref{surf01}(d). In both the dimensions, the maximum height of
the surfaces on regular lattice ($\phi=0$) are much higher than that
on random network ($\phi=1$). Though the maximum height is very less
in $1$d for $\phi=1$, all the lattice site toppled more than once
whereas in $2$d, the toppling surface on random network consists of
mostly singly toppled sites, only $0.06\%$ of the sites toppled more
than once. The toppling surfaces are found very different on regular
lattice and random network in different dimensions.

\begin{figure}[t]
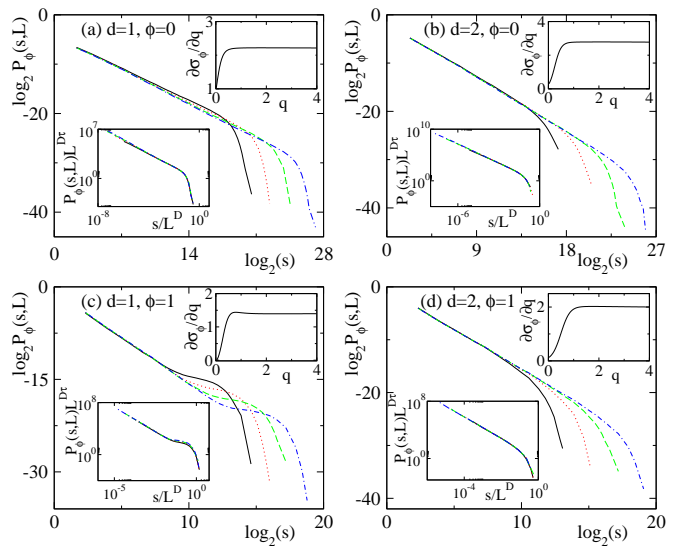

\centerline{\hfill 
  \psfig{file=figure_2a.eps,
    width=0.24\textwidth} \hfill 
  \psfig{file=figure_2b.eps,
    width=0.24\textwidth} \hfill}
\centerline{\hfill 
  \psfig{file=figure_2c.eps,
    width=0.24\textwidth} \hfill
  \psfig{file=figure_2d.eps,
    width=0.24\textwidth} \hfill}
\caption{(Color online) Plot of $P_\phi(s,L)$ of all the avalanches
  for $\phi=0$ (a) for $1$d and (b) for $2$d. $P_\phi(s,L)$ for
  $\phi=1$ are plotted in (c) and (d) for $1$d and $2$d
  respectively. Different curves corresponds to different system size
  $L$ as, for $1$d, $L=2^{10}$ (black solid line), $L=2^{11}$ (red
  dotted line), $L=2^{12}$ (green dashed line), and $L=2^{13}$ (blue
  dotted dashed line), whereas, for $2$d they are $L=2^{7}$ (black
  solid line), $L=2^{8}$ (red dotted line), $L=2^{9}$ (green dashed
  line), $L=2^{10}$ (blue dotted dashed line). Upper inset in each
  figure shows the plot of $\partial \sigma_{\phi}(q)/\partial q$
  against $q$ whereas, the lower inset shows the corresponding FSS
  data collapse.}
\label{Ps_all_1d-2d}
\end{figure}

\subsection{Moment analysis at $\phi = 0 \ \& \ 1$}
The critical steady state of sandpile model is mostly characterized by
power-law scaling of the probability distributions of avalanche size
($s$) occurring in the steady state. For a given $\phi$ and $L$, the
probability to have an avalanche of size $s$ is given by
$N_{s,\phi}/N_{\rm tot}$ where $N_{s,\phi}$ is the number of
avalanches of size $s$ out of total number of avalanches $N_{\rm tot}$
generated at the steady state. The distribution of $s$ follows a power
law scaling with a well defined exponent $\tau$ and obeys FSS
\cite{deMenechPRE98,*tebaldiPRL99}. The FSS form of the probability
distribution of $s$ in DSSM is given by
\begin{equation}
\label{psL}
P_\phi(s,L) = s^{-\tau}f_\phi\left[\frac{s}{L^{D}}\right]
\end{equation}
where $f_\phi$ is a $\phi$ dependent scaling function and $D$ is the
capacity dimension. Very often the power law scaling is found to
sustain over a short range of avalanche sizes and hinders precise
extraction of the exponent $\tau$ from the slope of the plot of
$P_\phi(s,L)$ against $s$ in double logarithmic scale. A more reliable
estimate of the exponent can be made employing moment analysis
\cite{karmakarPRL05,lubeckPRE00a}. For a given $\phi$, the $q$th
moment of $s$ is defined as
\begin{equation}
\label{sq}
\langle s^q(L) \rangle_\phi = \int_0^{\infty} s^{q}P_\phi(s,L)ds \sim
L^{\sigma_\phi(q)}
\end{equation}
where,
\begin{equation}
\sigma_\phi(q)=D(q-\tau+1)
\label{sigmaq}
\end{equation}
is the moment scaling function for $q>\tau-1$ (for $q<\tau-1$,
$\sigma_\phi(q)=0$). Values of $\sigma_\phi(q)$ are estimated from the
slope of the plots of $\langle s^q(L)\rangle_\phi$ versus $L$ in
double logarithmic scale for $400$ equidistant values of $q$ between
$0$ and $4$. The value of $D$ can be measured from the saturated value
of $\partial\sigma_\phi(q)/\partial q$ in large $q$ limit. The
derivative $\partial\sigma_\phi(q)/\partial q$ is determined
numerically by finite-difference method. Once $D$ is known the
exponent $\tau$ can be estimated from Eq. (\ref{sigmaq}) using the
value of $\sigma_\phi(1)$.

\noindent {\bf{\em All avalanches} :} $P_\phi(s,L)$ of all the
avalanches for various values of $L$ are presented in
Fig. \ref{Ps_all_1d-2d} for $1$d and $2$d for $\phi=0$ and
$1$. Reasonable power law scaling are observed for these extreme
values of $\phi$ in both the dimensions. The flat tail in
$P_\phi(s,L)$ for $\phi=1$ in $1$d is due to large dissipative
avalanches which will be discussed later separately. Employing moment
analysis, values of $D$ and $\tau$ are estimated at all four
situations. For $\phi=0$, estimates of $D$ are found to be
$2.21\pm0.02$ and $2.76\pm 0.02$ for $1$d and $2$d respectively. Since
for $\phi=0$, $\sigma_{\phi}(1)\approx 2$ in both the dimensions, the
values of $\tau$ estimated from Eq. (\ref{sigmaq}) are $1.09\pm 0.02$
in $1$d and $1.28\pm 0.01$ in $2$d. As expected, the exponents are
found very close to the reported values for SSM on regular lattice in
respective dimensions
\cite{dickmanPRE03,dickmanBJP00,huynhJSM11,*huynhPRE12}, for instance
in $1$d $\tau=1.112\pm 0.006$, $D=2.253 \pm 0.014$ and in $2$d
$\tau=1.273\pm 0.002$, $D=2.750 \pm 0.006$ for SSM. Whereas for
$\phi=1$, the values of $D$ are found to be $1.39\pm 0.02$ and
$\approx 2$ in $1$d and $2$d respectively. In $2$d, the avalanches on
random network ($\phi=1$) consist mostly single toppled nodes, hence
$D\approx 2$ is expected whereas the value of $D>1$ in $1$d suggests
that the avalanches consist of multiple toppled nodes. In both the
dimensions, the value of $\tau$ for $\phi=1$ is $\approx 1.50$, the
mean-field value as obtained in branching processes
\cite{christensenPRE93,bonabeau95,gohPRL03}. The values of the
exponents are listed in Table \ref{table2}. The FSS form of
$P_{\phi}(s,L)$ is verified by plotting the scaled distribution
$P_{\phi}(s,L)L^{D\tau}$ against the scaled variable $s/L^D$ in the
respective lower inset of Fig. \ref{Ps_all_1d-2d} using the respective
values of the critical exponents obtained.

\begin{figure}[t]
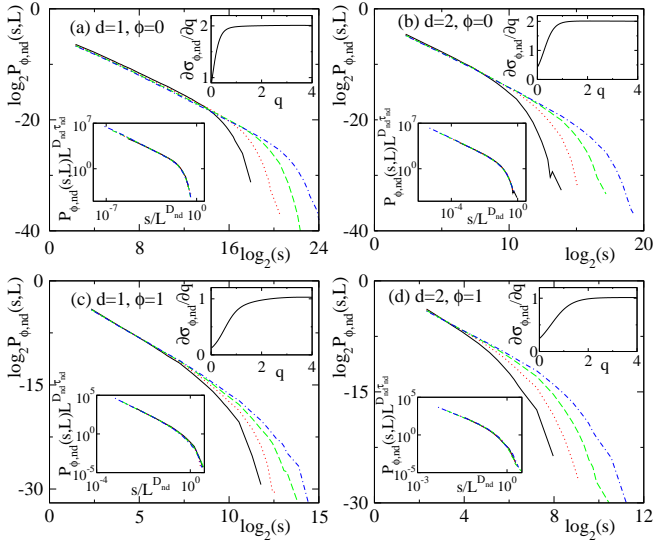

\centerline{\hfill 
  \psfig{file=figure_3a.eps,width=0.24\textwidth}\hfill 
  \psfig{file=figure_3b.eps,width=0.24\textwidth}\hfill
}
\centerline{\hfill 
  \psfig{file=figure_3c.eps,width=0.24\textwidth}\hfill
  \psfig{file=figure_3d.eps,width=0.24\textwidth}\hfill
}
\caption{(Color online) Plot of $P_{\phi,nd}(s,L)$ for different $L$
  (with same symbol as in Fig. \ref{Ps_all_1d-2d}) for $\phi=0$ in
  (a) for $1$d and in (b) for $2$d. For $\phi=1$ the same has been
  plotted in (c) and (d) for $1$d and $2$d respectively. Insets in
  each figure are same as that of Fig. \ref{Ps_all_1d-2d} but for
  non-dissipative avalanches.}
\label{Ps_nondiss_1d-2d}
\end{figure}

\noindent {\bf {\em Non-dissipative and dissipative avalanches}:}
Avalanches are now classified into non-dissipative and dissipative
avalanches. During the evolution, a dissipative avalanche must
dissipate at least a sand grain once whereas no sand grain be
dissipated in a non-dissipative avalanche. The avalanche size
distribution $P_\phi(s,L)$ can be written in terms of
$P_{\phi,nd}(s,L)$ and $P_{\phi,d}(s,L)$, the distributions of
non-dissipative and dissipative avalanches, as
\begin{equation}
P_\phi(s,L)=P_{\phi,nd}(s,L)+P_{\phi,d}(s,L)
\end{equation}
with
\begin{equation}
P_{\phi,nd}(s,L) = \frac{n_{s,nd}}{N_{tot}} \ \ \ {\rm and} \ \ \ 
P_{\phi,d}(s,L) = \frac{n_{s,d}}{N_{tot}}, 
\end{equation}
where $n_{s,nd}$ and $n_{s,d}$ are number of non-dissipative and
dissipative avalanches of size $s$ out of total $N_{tot}$
avalanches. First, the analysis of non-dissipative avalanches is given
and then that of dissipative avalanches is presented.

\begin{figure}[t]
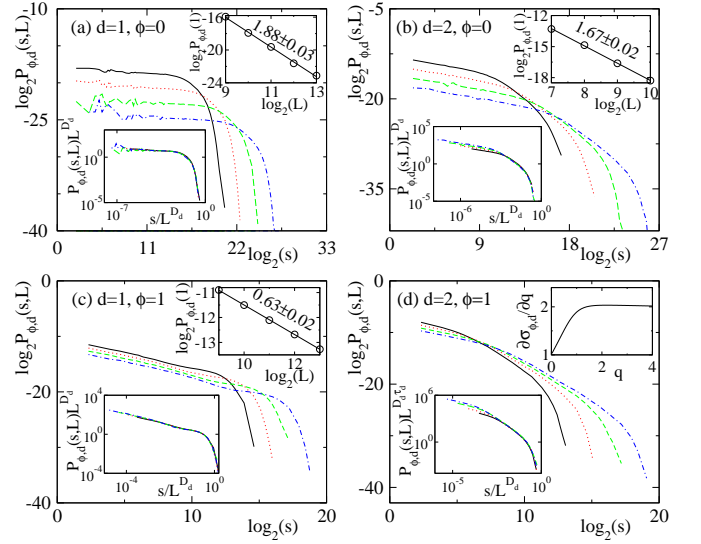

\centerline{\hfill 
  \psfig{file=figure_4a.eps,
    width=0.24\textwidth} \hfill 
  \psfig{file=figure_4b.eps,
    width=0.24\textwidth} \hfill}
\centerline{\hfill 
  \psfig{file=figure_4c.eps,
    width=0.24\textwidth} \hfill
  \psfig{file=figure_4d.eps,
    width=0.24\textwidth} \hfill}
\caption{(Color online) Plot of $P_{\phi,d}(s,L)$ for various $L$
  (with same symbol as in Fig. \ref{Ps_all_1d-2d}) for $\phi=0$ in
  (a) for $1$d and in (b) for $2$d. For $\phi=1$ the same has been
  plotted in (c) and (d) for $1$d and $2$d respectively. Upper insets:
  (a)-(c) shows the plot of $P_{\phi,d}(1,L)$ against $L$, (d) shows
  variation of $\partial \sigma_{\phi,d}(q)/\partial q$ against
  $q$. Lower insets shows the corresponding FSS data collapse (see
  text).}
\label{Ps_diss_1d-2d}
\end{figure}

The FSS form of the distribution $P_{\phi,nd}(s,L)$ is
assumed to be
\begin{equation}
\label{psLnd}
P_{\phi,nd}(s,L) = s^{-\tau_{nd}}f_{\phi,nd}\left[\frac{s}{L^{D_{nd}}}\right]
\end{equation}
where $f_{\phi,nd}$ is a scaling function, $\tau_{nd}$ and $D_{nd}$
are the respective exponents. $P_{\phi,nd}(s,L)$ for $\phi=0$ and $1$
are plotted in Fig. \ref{Ps_nondiss_1d-2d} for several values of $L$
for both $1$d and $2$d. Performing moment analysis, the values of
$D_{nd}$ are found as $D_{nd}\approx 2$ for $\phi=0$ and
$D_{nd}\approx 1$ for $\phi=1$ in both $1$d and $2$d. It could be
recalled here that the dissipation factor is chosen from the inverse
of $\langle n_\phi\rangle$. On an average the avalanche of size
$s>\langle n_\phi\rangle/2$ must dissipate at least one sand grain
(the factor $2$ is for one toppling consists two sand transfer). Since
$\langle n_\phi \rangle \sim L^2$ as $\phi \rightarrow 0$ due to
diffusive behaviour of random walker on regular lattice and $\langle
n_\phi \rangle \sim L$ as $\phi \rightarrow 1$ for super-diffusive
behaviour of random walker on random network \cite{bhaumikPRE13}, the
cutoff of $P_{\phi,nd}(s,L)$ must scales with $L$ in the same way as
$\langle n_\phi\rangle$ scales with $L$. Knowing the values of
$D_{nd}$ and $\sigma_{\phi,nd}(1)$, the values of $\tau_{nd}$ are
estimated. The values of $\sigma_{\phi,nd}(1)$ are found as $1.78$ for
$\phi=0$ and $0.48$ for $\phi=1$ in $1$d. Accordingly,
$\tau_{nd}=1.11(2)$ for $\phi=0$ and $\tau_{nd}=1.52(2)$ for $\phi=1$
in $1$d. The power law scaling of $P_{\phi,nd}(s,L)$ is found similar
to that of $P_{\phi}(s,L)$ as the values of $\tau$ and $\tau_{nd}$ are
found more or less same for both the distributions for $\phi=0$ and
$1$. Whereas in $2$d, the values of the exponents are found as
$\tau_{nd}=1.29\pm 0.01$ for $\phi=0$ (since
$\sigma_{\phi,nd}(1)=1.4$) and $\tau_{nd}=1.54\pm 0.05$ for $\phi=1$
as $\sigma_{\phi,nd}(1)=0.5$. On regular lattice it is the SSM result
whereas on random network it is the mean-field result. The values of
$D_{nd}$ and $\tau_{nd}$ for non-dissipative avalanches are listed in
Table \ref{table2}. Using the values of $\tau_{nd}$ and $D_{nd}$, a
reasonable data collapse is obtained for $P_{\phi,nd}(s,L)$ as shown
in the lower insets of Fig. \ref{Ps_nondiss_1d-2d}.

\begin{table}[t]
\centering
\begin{tabular}{c@{\hspace{.6cm}}c@{\hspace{.6cm}}c@{\hspace{.6cm}}c@{\hspace{.5cm}}c}
\hline\hline
 &&non-dissipative&dissipative&all\\
\cline{2-5}
\multicolumn{1}{c}{\multirow{4}{*}{$1$d}}&\multicolumn{1}{c}{\multirow{2}{*}{$\phi=0$}}&$D_{nd}=2.009$&$D_d=2.214$&$D=2.215$\\
&&$\tau_{nd}=1.11(2)$&$\tau_d=0.14(1)$&$\tau=1.09(2)$\\
\cline{2-5}
&\multicolumn{1}{c}{\multirow{2}{*}{$\phi=1$}}&$D_{nd}=1.022$&$D_d=1.402$&$D=1.395$\\
&&$\tau_{nd}=1.52(2)$&$\tau_d=0.54(2)$&$\tau=1.50(1)$\\
\hline\hline
\multicolumn{1}{c}{\multirow{4}{*}{$2$d}}&\multicolumn{1}{c}{\multirow{2}{*}{$\phi=0$}}&$D_{nd}=2.004$&$D_d=2.791$&$D=2.764$\\
&&$\tau_{nd}=1.29(1)$&$\tau_d=0.40(2)$&$\tau=1.28(1)$\\
\cline{2-5}
&\multicolumn{1}{c}{\multirow{2}{*}{$\phi=1$}}&$D_{nd}=1.013$&$D_d=2.023$&$D=2.008$\\
&&$\tau_{nd}=1.54(5)$&$\tau_d=1.45(3)$&$\tau=1.51(2)$\\
\hline\hline
\end{tabular}
\caption{\label{table2} Best estimated values of $D$ and $\tau$ for
  DSSM in $1$d and $2$d at $\phi=0$ and at $\phi=1$ for
  non-dissipative avalanches, dissipative avalanches, and all
  avalanches. As the values of $D$ is estimated from
  $\partial\sigma_\phi(q)/\partial q$ the error in determination of
  $D$ is $2\Delta q$ i.e. $\pm 0.020$. The number in the parentheses
  is the uncertainty of last digit of the value $\tau$ determined from
  the scaling relations.}
\end{table}

The size distribution of dissipative avalanches $P_{\phi,d}(s,L)$ for
several values of $L$ are presented in Fig. \ref{Ps_diss_1d-2d} for
$\phi=0$ and $1$ in both the dimensions. Interestingly, the
distributions $P_{\phi,d}(s,L)$ are very different in nature than the
corresponding $P_{\phi,nd}(s,L)$. Preliminary estimate of the size
distribution exponent $\tau_d$ by linear least square fit to the data
points in double-logarithmic scale reveals that $\tau_d<1$ except for
$\phi=1$ in $2$d. Following Christensen and co-workers
\cite{faridNJP06,*christensenEPJB08}, a new scaling form of
$P_{\phi,d}(s,L)$ is proposed as
\begin{equation}
\label{psL1}
P_{\phi,d}(s,L) =
s^{-\tau_d}L^{D_d(\tau_d-1)}f_{\phi,d}\left[\frac{s}{L^{D_d}}\right]
\end{equation}
where $f_{\phi,d}$ is a new scaling function, $\tau_d$ and $D_d$ are
exponents for dissipative avalanches. The moment of such a
distribution is obtained as $\langle s^q(L) \rangle_{\phi,d} \sim
L^{\sigma_{\phi,d}(q)}$ where $\sigma_{\phi,d}(q)=qD_d$. Noticeably,
the moment exponent $\sigma_{\phi,d}(q)$ becomes independent of the
size distribution exponent $\tau_d$. Performing moment analysis for
both $1$d and $2$d, the values of $D_d$ for dissipative avalanches are
found close to that of the all avalanches as presented in Table
\ref{table2}. In the limit $s/L^{D_d}\rightarrow 0$, the scaling
function $f_{\phi,d}[s/L^{D_d}]$ becomes a constant and the
distribution is given by $P_{\phi,d}(s,L) \approx
s^{-\tau_d}L^{D_d(\tau_d-1)}$. Consequently, $P_{\phi,d}(1,L) \sim
L^{D_d(\tau_d-1)}$ for $s=1$. The exponent $\tau_d$ is then estimated
from the slope of the plot of $P_{\phi,d}(1,L)$ vs $L$ in double
logarithmic scale as presented in the upper insets of
Figs. \ref{Ps_diss_1d-2d} (a), \ref{Ps_diss_1d-2d}(b), and
\ref{Ps_diss_1d-2d}(c). The values of $\tau_d$ are estimated as
$0.14\pm 0.01$ for $\phi=0$ and $0.54 \pm 0.01$ for $\phi=1$ in
$1$d. It is interesting to note that such a flat distribution (i.e
$\tau_d \rightarrow 0$ for $\phi=0$ in $1$d) is also reported by
Amaral and Lauritsen \cite{amaralPRE96} for the dissipative avalanches
of $1$d rice pile model. In contrary to the present observation,
Dickman and Campelo \cite{dickmanPRE03} found a power law scaling of
$P_d(s,L)$ with exponent $\tau_d=0.637$ for SSM with boundary
dissipation on $1$d regular lattice. In $2$d, the exponent $\tau_d$ is
obtained here as $0.40 \pm 0.02$ for $\phi=0$ whereas Dickman and
Campelo \cite{dickmanPRE03} reported $\tau_d=0.98$ for SSM on $2$d
regular lattice with boundary dissipation. Thus the scaling behaviour
of dissipative avalanches of DSSM is very different from that of
dissipative avalanches of SSM with boundary dissipation in both the
dimensions at $\phi=0$. Such difference in the scaling behaviour for
dissipative avalanches with different modes of dissipation is probably
due to different topological properties of network in the bulk and at
the boundary because the degree of a node at the boundary is different
from that of a node in the bulk. Moreover, it should be noted that for
the model of boundary dissipation, Dickman and Campelo introduced a
logarithmic correction in the distribution of dissipative avalanches
and the distribution was given by
\begin{figure}[t]
\centerline{\hfill
  \psfig{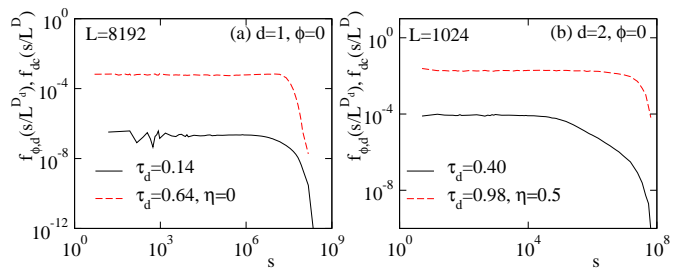}\hfill
}
\caption{(Color online) Plot of $f_{\phi,d}(=P_{\phi,d}(s)s^{\tau_d})$
  (in solid black line) and
  $f_{dc}(=P_{\phi,d}(s)s^{\tau_d}[\mbox{ln}(s)]^{-\eta})$ (in dashed
  red line) against $s$ for $1$d, $L=8192$, in (a) and for $2$d,
  $L=1024$, in (b). The values of the $\tau_d$ and $\eta$ for $f_{dc}$
  are taken from Ref. \cite{dickmanPRE03}. Since the plots are for a
  given $L$, the $L$ dependency of the argument is dropped.}
\label{fsLD}
\end{figure}
\begin{equation}
\label{psdc}
P_{dc}(s,L)
=s^{-\tau_d}[\mbox{ln}(s)]^{\eta}f_{dc}\left[\frac{s}{L^{D}}\right]
\end{equation}
where $\eta$ is an another exponent. In order to verify whether such
correction to scaling is present in the present model with bulk
dissipation or not, the scaling function $f_{\phi,d}$ given in
Eq. (\ref{psL1}) for $\phi=0$ is plotted in Figs. \ref{fsLD}(a) and
\ref{fsLD}(b) for $1$d and $2$d respectively. For comparison, the
scaling function $f_{dc}$ that of the model with boundary dissipation
given in Eq. (\ref{psdc}) is also plotted in the respective plots. It
can be seen that without any correction, the scaling function
$f_{\phi,d}$ is reasonably constant over a wide range of $s$ in double
logarithmic scale in the case of bulk dissipation whereas that
requires a correction to scaling, $[\mbox{ln}(s)]^{\eta}$, in the case
of boundary dissipation for $d=2$ ($\eta=0.5$) as observed by Dickman
and Campelo \cite{dickmanPRE03}. Hence the scaling forms considered
here for the model with bulk dissipation are not subject to any
logarithmic correction. However, the scaling behaviour of all
avalanches are found to be same for both the models as reported in ref
\cite{malcaiPRE06}. This is because the leading singularity is
provided by non-dissipative avalanches. In order to verify the form of
the scaling function given the Eq. (\ref{psL1}), data collapse has
been performed by plotting $P_{\phi,d}(s,L)L^{D_d}$ against
$s/L^{D_d}$ for different values of $L$. Reasonable data collapse for
different $P_{\phi,d}(s,L)$ are obtained as shown in the respective
lower insets of Figs. \ref{Ps_diss_1d-2d}. For $\phi=1$ in $2$d, the
FSS form of the distribution $P_{\phi,d}(s,L)$ is expected to follow
the usual distribution as given in Eq. (\ref{psL}). From the plot of
$\partial\sigma_{\phi,d}(q)/\partial q$ vs $q$ as given in upper inset
of Fig. \ref{Ps_diss_1d-2d}(d), $D_d$ is found to be $2.023\pm 0.020$,
again close to the value of all avalanches. The value of $\tau_d$ is
estimated from Eq. (\ref{sigmaq}) as $1.45\pm 0.03$ little less than
mean-field result as obtained for the all avalanches. However, taking
$\tau_d=3/2$ and $D_d=2.023$ the best data collapse is obtained, given
in lower inset of Fig. \ref{Ps_diss_1d-2d}(d), which confirms the
respective form of the scaling function. It should be noted here that
the value of $D_d$ for dissipative avalanches are very close to the
value of $D$ of the all avalanches for both the extreme values of
$\phi$ in both the dimensions because the large avalanches which are
responsible for cutoff of the distribution of all avalanches are
mostly dissipative, and in the moment analysis the leading
contribution comes from those large dissipative avalanches.

\subsection{Small world regime}
\begin{figure}[t]
\centerline{\hfill
  \psfig{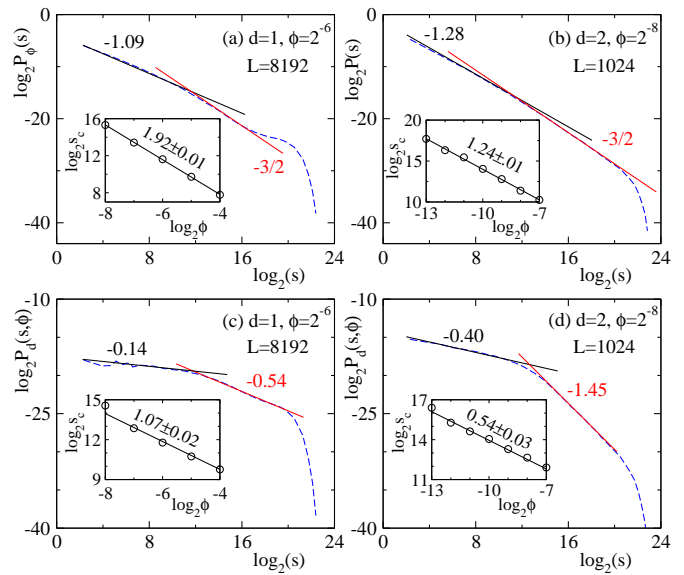}\hfill
}
\caption{(Color online) Plot of $P_\phi(s)$ against $s$ in (a) for
  $1$d with $\phi=2^{-6}$ for $L=8192$ and in (b) for $2$d with
  $2^{-8}$ for $L=1024$. The corresponding $P_d(s,\phi)$ is shown in
  (c) for $1$d and in (d) for $2$d. The solid lines with required
  slope through data points are guide to eye. The variation of $s_c$
  with $\phi$ is shown in respective inset. The error in estimation
  of $s_c$ is of the order of symbol size.}
\label{Ps-swn}
\end{figure}
\noindent {\bf {\em Scaling properties}:} Since SWN preserves both the
characteristic of regular lattice and random network, it is important
to study the critical properties of the avalanche size distribution in
the SWN regime, $2^{-12}<\phi <0.1$. The size distribution
$P_{\phi}(s)$ of all the avalanches are plotted in
Fig. \ref{Ps-swn}(a) for $1$d and in Fig. \ref{Ps-swn}(b) for $2$d. In
Fig. \ref{Ps-swn}(c) and (d), $P_{\phi,d}(s)$ are plotted for $1$d and
$2$d respectively. For $1$d and $2$d, the values of $\phi$ used are
$2^{-6}$ and $2^{-8}$ respectively for both the
distributions. Interestingly, both the distributions $P_{\phi}(s)$ and
$P_{\phi,d}(s)$ exhibit their respective scaling forms on regular
lattice ($\phi=0$) and random network ($\phi=1$) in the same
distribution. The straight lines with respective slopes in these plots
are guide to eye. The crossover from one scaling form to other occurs
at their respective crossover avalanche size $s_c$ for $P_{\phi}(s)$
and $P_{\phi,d}(s)$. For $s<s_c$, the avalanches are small, compact
and mostly confined on regular lattice whereas for $s>s_c$, they are
large, sparse and mostly exposed to random network. Since
$P_{\phi,nd}(s)$ and $P_{\phi}(s)$ have similar scaling behaviour,
$P_{\phi,nd}(s)$ display a similar crossover scaling as that of
$P_{\phi}(s)$. The coexistence of more than one scaling forms in the
same distribution of avalanche properties for different sandpile model
are already reported in the literature
\cite{bhaumikPRE13,lahtinenPHYA05,hooreJPA13,*araghiPRE15,*moosaviPRE15}. The
crossover scaling is found to occur for a wide range of $\phi$ within
SWN regime for both $P_{\phi}(s)$ and $P_{\phi,d}(s)$. As one expects
the scaling form of regular lattice as $\phi\rightarrow 0$ and that of
random network as $\phi\rightarrow 1$, the value of $s_c$ is found to
depend on $\phi$ for both the distributions. The dependence of $s_c$
on $\phi$ is assumed as
\begin{equation}
s_c \sim \phi^{-\alpha}
\label{sc}
\end{equation}
where $\alpha$ is an exponent. The value of $\alpha$ for all
avalanches can be obtained by simple arguments. From the conditional
expectation of avalanche size for a fixed avalanche area, one expects
$s_c \sim a_c^{\gamma_{sa}}\approx \xi^{d\gamma_{sa}}$, where $a_c$ is
the average avalanche area for the avalanches of size $s_c$,
$\gamma_{sa}$ is an exponent \cite{christensenPRE93}, and $\xi$ is the
crossover length scale below which SWN behaves as regular lattice
\cite{newmanPRE99,*newmanPLA99,mendesEPL00}. As $\xi \sim
\phi^{-1/d}$, one obtains $s_c\sim \phi^{-\gamma_{sa}}$ and has
$\alpha=\gamma_{sa}$. However, a dissipative avalanche occurs only
after a required number of toppling equivalently $\langle n_{\phi}
\rangle$. As $\langle n_{\phi} \rangle \sim \phi^{-1/d}$ in the large
$\phi$ limit corresponding to random network, one expects
$s_{c}\sim\phi^{-1/d}$ with $\alpha=1/d$. For a given $\phi$ the value
of $s_c$ is estimated from the intersection point of the straight
lines with required slope in the respective regions. The estimated
values of $s_c$ is then plotted against $\phi$ in double logarithmic
scale in the respective insets of Fig. \ref{Ps-swn}. It can be seen
that in all cases $s_c$ shows a reasonable power-law scaling with
$\phi$. By linear least square fit through the data points the values
of $\alpha$ for all avalanches are found to be $1.92\pm 0.01$ for $1$d
and $1.24\pm 0.01$ for $2$d which are very close to the $\gamma_{sa}$
values at $\phi=0$ in both the dimensions
\cite{nakanishiPRE97,benhurPRE96,*santraPRE07}. On the other hand, for
dissipative avalanches it is found that $\alpha=1.07\pm 0.02$ for $1$d
and $\alpha=0.54\pm 0.03$ for $2$d again close the inverse of
respective dimensions.

\begin{figure}[t!]
\centerline{\hfill 
  \psfig{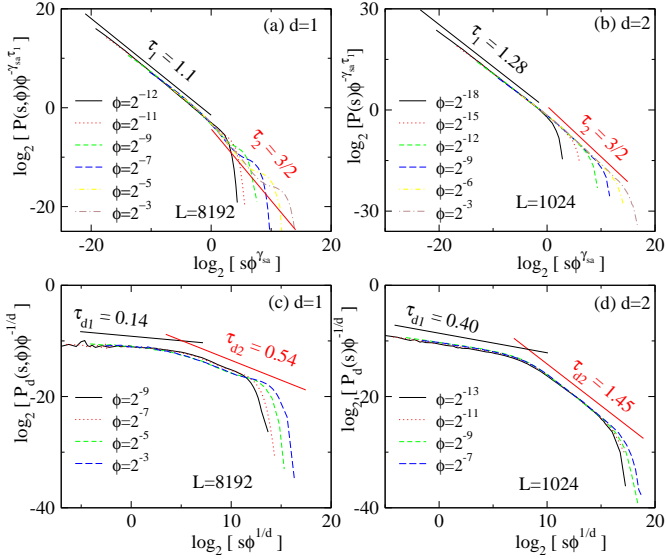}\hfill}
\caption{(Color online) Scaled probability distribution (for all
  avalanches) against scaled variable is plotted for selected values
  of $\phi$ in (a) for $1$d with $L=8192$ and in (b) for $2$d with
  $L=1024$ to verify the scaling forms given in
  Eq. (\ref{scaling3}). The same has been verified for dissipative
  avalanches in (c) for $1$d and in (d) for $2$d.}
\label{SC1}
\end{figure}
\noindent {\bf {\em Coexistence scaling}:} Since FSS forms of
$P_{\phi}(s,L)$ and $P_{\phi,d}(s,L)$ are found to be satisfied both
on regular lattice and on random network, they should also be
satisfied on SWN. Instead of FSS form, the $\phi$ dependence of these
distributions are then verified on SWN for a fixed $L$. A generalized
scaling form for $P(s,\phi)$ for the all avalanches on SWN is proposed
as
\begin{equation}
\label{scaling}
P(s,\phi) = \left \{ \begin{array}{ll} s^{-\tau_{1}} {\sf
    f}\left(\frac{s}{s_c(\phi)}\right) & \mbox{for $s \leqslant s_c$}
  \\ s^{-\tau_{2}} {\sf
    g}\left(\frac{s}{s_c(\phi)}\right) & \mbox{for
    $s \geqslant s_c$ }
       \end{array} \right. 
\end{equation}
where ${\sf f}$ and ${\sf g}$ are the respective scaling functions and
$\tau_{1}$, $\tau_{2}$ are the corresponding critical exponents in
$s<s_c$ and $s>s_c$ regions respectively. At $s=s_c$ for a given
$\phi$, the limiting values of $P(s,\phi)$ from both the regions must
be same. As $s_c\sim \phi^{-\alpha}$, then one should have
$\phi^{\tau_{1}\alpha}{\sf f}(1) = \phi^{\tau_{2}\alpha}{\sf g}(1)$.
Hence, the $\phi$ independent scaled distribution can be obtained as
\begin{eqnarray}
\label{scaling3}
P(s,\phi)\phi^{-\alpha\tau_{1}} &=& \left \{ 
\begin{array}{ll} 
  \left(s\phi^\alpha\right)^{-\tau_{1}} {\sf
    f}(s\phi^\alpha) & \mbox{for $s\leqslant s_c$}
  \\ \left(s\phi^\alpha\right)^{-\tau_{2}} {\sf
    f}(s\phi^\alpha) &\mbox{for $s\geqslant s_c$ }\\
\end{array} \right. 
\end{eqnarray}
in terms of a single scaling function ${\sf f}$
\cite{bhaumikPRE13}. Such scaling form is also found to exist in the
dynamic scaling of roughness of fractured surfaces
\cite{lopezPRE98,*morelPRE98}. To verify the scaling forms given in
Eq. (\ref{scaling3}), the scaled probabilities for all avalanches are
plotted against the scaled variable in Figs. \ref{SC1}(a) and
\ref{SC1}(b) for $1$d and $2$d respectively taking
$\alpha=\gamma_{sa}$. It can be seen that a good data collapse is
obtained using $\gamma_{sa}=2$, $\tau_{1}=1.1$ for $1$d and using
$\gamma_{sa}=1.26$, $\tau_{2}=1.28$ for $2$d. The straight lines with
required slopes in the respective regions are guide to eye. It
confirms the validity of the proposed scaling function form given in
Eq. (\ref{scaling}). Similarly, a generalized size distribution
function can be written for dissipative avalanches around its
crossover size $s_{c}$ taking $\alpha=1/d$. The scaled probabilities
for $P_d(s,\phi)\phi^{-1/d}$ for dissipative avalanches are plotted
against the scaled variable $s\phi^{1/d}$ in Figs. \ref{SC1}(c) and
\ref{SC1}(d) for $1$d and $2$d respectively. Reasonable data collapse
is obtained as expected. It is then important to notice that if a
dynamical model like sandpile is studied on SWN, multiple scaling
forms of an event size will coexist in the distribution of the same.

\section{Summary and Conclusion} 
A dissipative stochastic sandpile model is developed and its critical
properties are studied on SWNs both in $1$d and $2$d for a wide range
of shortcut density $\phi$. The non-dissipative avalanches display
usual stochastic scaling of SSM on regular lattice ($\phi=0$) and
mean-field scaling on random network ($\phi=1$) as that of all
avalanches. However, the dissipative avalanches represent a number of
novel scaling properties on regular lattice as well as on random
network in both $1$d and $2$d. The scaling behaviour of these
avalanches on regular lattice is found to be very different from
Dickman-Campelo scaling as observed with open boundary in both the
dimensions. The bulk dissipation is found to have non-trivial effect
on dissipative avalanches over the boundary dissipation. No
logarithmic correction to scaling is found to occur as it was required
for these avalanches on regular lattice with boundary dissipation. A
set of new scaling exponents are found to describe the scaling of
dissipative avalanches on regular lattice and random network. On SWN,
in the intermediate range of $\phi$, the model exhibits coexistence of
more than one scaling forms both in $1$d and $2$d around a crossover
size $s_c(\phi)$. For non-dissipative and dissipative avalanches,
however, the crossover size $s_c$ scales with $\phi$ with two
different exponents. The small, compact avalanches of size $s<s_c$
mostly confined on regular lattice are found to obey the usual SSM
scaling whereas the large, sparse avalanches of size $s>s_c$ exposed
to random network are found to obey mean-field scaling. A coexistence
scaling form of the avalanche size distribution function around $s_c$
is proposed and numerically verified. Therefore, SWN can be considered
as a segregator of several scaling forms that appear in the event size
distribution in a dynamical system.

\bigskip
\noindent{\bf Acknowledgments:} This work is partially supported by
DST, Government of India through project No. SR/S2/CMP-61/2008. HB
thanks MHRD, Government of India for financial
assistance. Availability of computational facility, ``Newton HPC''
under DST-FIST project Government of India, of Department of Physics,
IIT Guwahati is gratefully acknowledged.

\bibliography{paper_revised.bbl}
\end{document}